\documentclass{article}

\usepackage{arxiv}

\usepackage[utf8]{inputenc} 
\usepackage[T1]{fontenc}    
\usepackage{hyperref}       
\usepackage{url}            
\usepackage{booktabs}       
\usepackage{amsfonts}       
\usepackage{nicefrac}       
\usepackage{microtype}      
\usepackage{lipsum}
\usepackage{graphicx}
\usepackage{amsmath}
\graphicspath{ {./images/} }

\title{Design and optimization of dihedral angle offsets for the next generation lunar retro-reflectors}

\author{
  Chensheng Wu \\
  Department of Electrical and Computer Engineering\\
  University of Maryland College Park\\
  College Park, MD 20742 \\
  \texttt{cwu2011@umd.edu} \\
   \And
 Douglas Currie \\
  Department of Physics\\
  University of Maryland College Park\\
  College Park, MD 20742 \\
  \texttt{currie@umd.edu} \\
   \AND
  Dennis Wellnitz \\
  Department of Astronomy\\
  University of Maryland College Park\\
  College Park, MD 20742 \\
   \texttt{wellnitz@umd.edu} \\
   \And
  Bradford Behr \\
  Department of Physics\\
  University of Maryland College Park\\
  College Park, MD 20742 \\
  \texttt{bradfordbehr@gmail.com} \\
  
}

\begin{document}
\maketitle

\begin{abstract}
Lunar laser ranging (LLR) to the Apollo retro-reflectors, which features the most long-lasting experiment in testing General Relativity theories, has remained operational over the past four decades. To date, with significant improvement of ground observatory conditions, the bottleneck of LLR accuracy lies in the retro-reflectors. A new generation of large aperture retro-reflectors with intended dihedral angle offsets have been suggested and implemented based on NASA's recent lunar projects to reduce its ranging uncertainty to be less than 1.0 mm. The technique relies on the retro-reflector's ability to offset its relative angular velocity with regard to a ground LLR observatory (LLRO), so that the LLR accuracy can be ensured along with the larger area of beam reflection. In deployment, solid corner-cube reflectors (CCRs) based on empirical successes of the Apollo 11 and 15 arrays have been selected for the next generation lunar reflectors (NGLRs) due to their stability against heat and dust problems on the Moon. In this work, we present the optical effects in designing the new retro-reflectors given various sets of intended diheral angle offsets (DAOs), and support the design principles with the measurements of of two manufactured NGLRs.
\end{abstract}

\keywords{Lunar laser ranging \and dihedral angle offset \and NGLR}

\section{Introduction}
The main objective of designing and deploying newer retro-reflectors on the Moon for the \textit{$21^{st}$} century LLR experiments is to improve both the photon return rate and accuracy for multiple ground stations \cite{viswanathan2020extending,murphy2008apache, currie2011lunar}. So far, estimated dust accumulation on the Apollo 15 retro-reflector array has reduced the effective number of retro-reflectors from 300 to around 30. Such degradation has been well compensated through increased laser power in the LLR over the past decades, so that an optimized steady photon return rate around 0.1 can be maintained. However, the practice will lose its effectiveness when more units in the Apollo 15 array get degraded by lunar dust, risking the continuity of the LLR experiments. Higher precision levels are also desired for future LLR, requiring the reflected photons to be more coherent returns than independent returns. Overall, the success of simultaneous and multi-point high-precision measurements of Earth-Moon distance and its dynamics from time to time offer assurance to test the fundamental truth of time-space law predicted by Einstein's general relativity theorem, which is viewed as one of the greatest experiments in human history \cite{williams2004progress, williams1976new, dickey1994lunar}. Retro-reflectors delivered through previous space exploration have well served the purpose of establishing the baseline of Earth-Moon distance and monitoring its changes. Recently, scientists came up with more ambitious foresight that many fundamental theories about the universe can be examined with more delicate reflectors on the Moon that reduces the measurement error to 1.0 mm or less \cite{murphy:lunar, martini:moonlight, dell2012probing}. In addition, ground observatories have already perfected their detection techniques over many years of improvement with the previous retro-reflectors on the Moon \cite{murphy2012apollo, murphy2000apache}. High precision ground measurements of gravitational waves, known as "LIGO", also achieved unhistorical success in detection \cite{abramovici1992ligo,harry2010advanced}. These scientific progress raises tremendous curiosities in using different approaches that can mutually confirm the findings and study them at deeper levels \cite{klimenko2016method,williams2009lunar,ciocci2017performance}. Currently, the photon return rate in Lunar laser ranging is approximately 0.1, meaning an average reception of one photon return per 10 pulses sent to the Moon. The significant loss is caused by diffraction loss over the great propagation distance, relative surface motion between the two astronomical bodies, and atmospheric turbulence near the surface of the Earth.   

Intuitively, photon return rate can be much improved with retro-reflectors of larger apertures, as a universal method to narrow the diffraction angle in beam spreading. Our new retro-reflector is designed to have an aperture diameter of 100 mm that is sufficient to elevate the central photon return by two magnitudes when compared with individual reflectors deployed on the Moon. The new accuracy challenge requires the retro to be more resilient to vibration and temperature changes, and the photon paths in the round trip be tractable \cite{dell2011creation,xie2010post}. In other words, the new retro needs to ensure minimum aberrations for incident photons at different times and locations through its aperture. These problems can be largely solved through a wise selection of materials and housing designs \cite{ciocci2017performance,currie2011lunar}. Therefore, a single unit of large aperture retro-reflector is expected to achieve the same or better results in LLR than previous retro-reflector arrays deployed on the Moon. However, a special problem occurs when the retro-reflector's aperture diameter increases to 100 mm, where the same observatory that sends out the pulses falls closer to the first diffraction zero than the preferred center of the reflected beam. This is caused by the relative surface motion between Earth and Moon during the approximated 2.5 seconds light travel time. One straightforward solution is to implement intentional dihedral angle offsets \cite{otsubo2011asymmetric} on the new retro-reflector to offset the relative motion. However, unlike a hollow corner-cube-reflector (CCR) that is easy to calculate the offset angles, a lunar retro-reflector needs to be dielectric to survive better under lunar dust, minimize heat absorption, endure the launch, and stay functional for decades of years. Consequently, it requires detailed and accurate analysis to track photon paths, account for depolarization at each total internal reflection (TIR), as well as additional phase error map (PEM) in manufacturing the modified retro-reflectors. The azimuth rotation of the new retro's aperture facing Earth upon the one-time deployment, on the other hand, introduces extra complication to the modification design with uncertain misalignment.

In this work, we demonstrate the analysis of beam reflection from a dielectric retro-reflector with intended dihedral angle offsets, and find the optimized designs for the next generation lunar retro-reflectors (NGLR) through numerical simulations \cite{currie2020next}. To the best of the authors' knowledge, this is the first time that such analysis has been done. Evaluation of two implemented retro-reflectors are also provided to determine their fitness for the \textit{$21^{st}$} century LLR tasks.

\section{Modified Retro-reflector with Dihedral Angle Offsets}

The idea of dihedral angle offsets was raised by Toshimichi and his colleagues for a large-aperture retro-reflector to offsets the velocity difference between the reflector and the ranging station. The optimized angle offsets can be calculated straightforwardly under the circumstances of a hollow retro-reflector. And the maximum error in photon travel distances per 1.0 arcecond angle offset per 1.0 meter spacing through the retro-reflection is well under \textit{$10\, \mu\!m$}, which is trivial comparing with the \textit{$1.0\, mm$} accuracy requirement. Exceptional work of such concept includes He's design of a 170 mm diameter hollow CCR for future LLR experiments \cite{he2018development}. On the other hand, as dust prevention and temperature treatment is critical for the NGLRs, researchers found it practical to choose solid dielectric retro-reflectors to survive the flight to the Moon and long-term LLR measurements. This in turn requires more complex modeling and analysis of the depolarization effects in the retro's TIRs, azimuth orientation in facing the Earth, as well as heat induced surface aberrations. In these regards, we provide a fundamental geometric and wave-optic integrated analysis to find the optimized dihedral angle offsets that needs to be implemented in a dielectric retro-reflector for future LLR tasks. Effects such as the higher order aberrations, the orientation of the retro-reflector facing the Earth, as well as input beam polarization are also integrated with the simulation for comprehensiveness.

\begin{figure}
\centering
\includegraphics[width=0.70\linewidth]{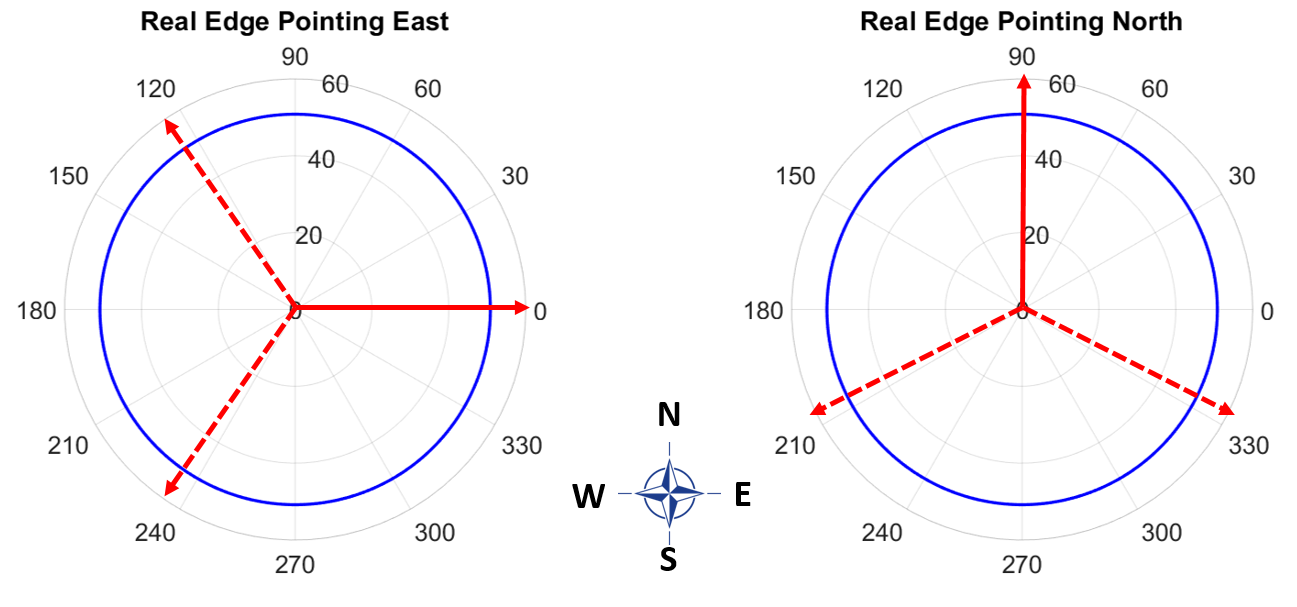}
\caption{Illustration of the real edge's azimuth orientations with corresponding dihedral angle offsets in scenarios of REE: real-edge pointing East (left), and REN: real-edge pointing North (right).}
\label{fig:Fig_1}
\end{figure}

Without loss of generality, we establish the coordinates based on the Earth's reference (or equivalently, the directions defined by looking from Moon to Earth) with \textit{$+x$} axis pointing East and \textit{$+y$} axis pointing North, respectively. The 3 real edges of the retro-reflector can be projected to the x-y plane to denote their pointing directions, assumed that the origin coincides with the apex point of the retro-reflector. Due to symmetry, there are two basic cases for the retro-reflector's azimuth orientation: real-edge-East (REE) where one real edge's projection in x-y plane lines up with the \textit{$+x$} axis, and real-edge-North (REN) where one real edge's projection in x-y plane lines up with the \textit{$+y$} axis. Correspondingly, the intended dihedral angle offset requires the REE case to tilt the back face opposing to the East pointing real edge, and the REN case to tilt the two adjacent back faces that intersect at the North pointing real edge, respectively. For alignments other than the two major cases of orientations, they are regarded as misalignment that will subdue the effectiveness of the modification. The misalignment could happen at physical deployment due to complex reasons, which will be briefly discussed in section 5 for real products. The 2D projected diagram of the two cases can be shown in Fig. \ref{fig:Fig_1}. 

In Fig. \ref{fig:Fig_1}, we use the blue circles to indicate the circular aperture of the retro-reflector that is 100 mm in diameter. The red solid arrow indicates the real-edge (that is marked to track the orientation of the retro-reflector), and the red dashed arrows indicate the other two edges after the intended modifications. For convenience of illustration, we use the name of "alignment edge" for both the East-pointing real edge in the REE case, and the North-pointing real edge in the REN case.

Based on the drawing, a heuristic estimation regarding the optimized dihedral angle offsets for the both the REE and the REN cases can be achieved using geometric optics. We later uses such estimation baseline to search for an optimized design by considering TIR depolarization and mutual interference in the reflected patterns, which is achieved through hybrid simulation using both geometric and wave optics.

We define the azimuth angle offsets in the projected x-y plane (shown in Fig. \ref{fig:Fig_1}) as $\delta$ for the two non-alignment edges. The value of $\delta$ is positive if the offsets turns counter-clockwise, and negative for clockwise turns. Correspondingly, their elevation angle offsets are denoted as $\Delta\alpha$ in the REE case and $\Delta\beta$ in the REN case. We have:
\begin{equation}
\Delta\alpha = -\sqrt{\frac{2}{3}}\delta.
\label{eq:Eq1}
\end{equation}

\begin{equation}
\Delta\beta = \sqrt{\frac{2}{3}}\delta.
\label{eq:Eq2}
\end{equation}

We have applied the \textit{$1^{st}$} order perturbation in deriving Eqs. (\ref{eq:Eq1}-\ref{eq:Eq2}), as the angle offsets are much less than unity (in the order of $10^{-6}$ rad). In both Eq. (\ref{eq:Eq1}) and Eq. (\ref{eq:Eq2}), positive values indicate the edge tilting away from the x-y plane and vice versa. Correspondingly, changes to the relative surface angles caused by the offsets are determined as:

\begin{equation}
\Delta\Phi_{REE} = \frac{4\sqrt{3}}{9}\delta.
\label{eq:Eq3}
\end{equation}

\begin{equation}
\Delta\Phi_{REN} = -\frac{4\sqrt{3}}{3}\delta.
\label{eq:Eq4}
\end{equation}

\begin{figure}[tbp]
\centering
\includegraphics[width=0.70\linewidth]{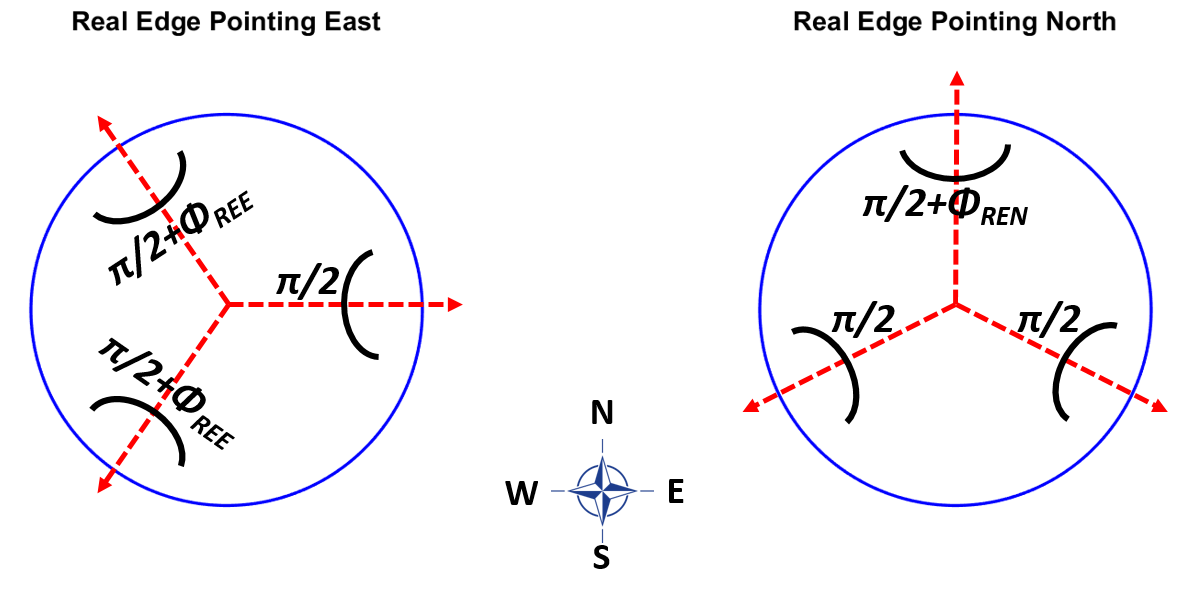}
\caption{Adjusted surface angles in scenarios of REE: real-edge pointing East (left), and REN: real-edge pointing North (right).}
\label{fig:Fig_2}
\end{figure}

In Eqs. (\ref{eq:Eq3}-\ref{eq:Eq4}) the pair-wise surface angle offsets referenced to a perfect retro-reflector are marked as \textit{$\Delta\Phi_{REE}$} and \textit{$\Delta\Phi_{REN}$} for the two major orientation scenarios, which are shown in Fig. \ref{fig:Fig_2}. In manufacturing, \textit{$\Delta\Phi_{REE}$} and \textit{$\Delta\Phi_{REN}$} can be measured by an interferometer with reasonable accuracy.

In each orientation scenario, the retro-reflector is deterministic given the surface angle offsets. The input waveform is also deterministic, because the Earth-Moon distance is about 384,400 km, and the wavefront of the incident pulse received by the retro-reflector over the long range diffraction (with turbulence influence considered) can be safely assumed to be a plane wave with normal incident angle. Therefore, in the retro-reflector design, we only need to consider the changes in the far-field-diffraction-pattern (FFDP) based on an incident planewave that fills the aperture. Furthermore, geometric optics can be used to investigate the TIRs and depolarization process, with negligible loss of accuracy when compared with wave optics. In other words, we trace the reflection inside the retro-reflector with geometric optics, and convert the output into a waveform to study its FFDP. The interim loss between geometric and wave optics inside the retro is a second order effect in ray distribution when compared with the magnitude of dihedral angle offsets (that represent the first order perturbation), which can be safely ignored in our analysis. Without loss of generality, we express the far-field diffraction in angular space, which is equivalent to the actual diffraction geometries divided by the propagation distance.         

\section{Method of NGLR reflection and far-field diffraction analysis}

In our geometric simulation, we employ a 10-D plenoptic function \cite{wu2015determining, mcmillan1995plenoptic,adelson1991plenoptic} for each ray in a high density ray tracing analysis to study the retro-reflector's reflection properties. The plenoptic function is marked as \textit{$P_n(x,y,z,E_s^x,E_s^y,E_s^z,E_p^x,E_p^y,E_p^z,\phi)$}, with \textit{$(x,y,z)$} tracking geometry of the ray's incident points at TIRs. Parameters \textit{$(E_s^x,E_s^y,E_s^z)$} and \textit{$(E_p^x,E_p^y,E_p^z)$} track polarization changes per internal reflection for s-wave and p-wave, respectively. And \textit{$\phi$} tracks phase changes in propagation. Intuitively, each ray can be understood as a regional segment of a plane wave that follows the law of Fresnel refraction. It is noticeable that the 10 parameters in the plenoptic function are not fully independent (such as the fact that s-p waves are orthogonal to each other), and many rays share identical parameters (such as the group of rays reflected by the same TIR sequence). However, because the surfaces of the modified retro-reflector may subject to minor aberrations caused by heat, dust and damage problems over many years of usage, we keep the redundancy in the ray tracing algorithm for future integration of various analysis. The depolarization effects at TIR follow the laws as \cite{arwin2004total,peck1962polarization,imbert1972calculation}:

\begin{equation}
\begin{split}
r_s = \frac{n_0\,cos\,\phi_0+jn_1\big[ \big(\frac{sin\,\phi_0}{sin\,\phi_c}\big)^2-1\big]^{1/2}}{n_0\,cos\,\phi_0-jn_1\big[ \big(\frac{sin\,\phi_0}{sin\,\phi_c}\big)^2-1\big]^{1/2}},\\
r_p = \frac{n_1\,cos\,\phi_0+jn_0\big[ \big(\frac{sin\,\phi_0}{sin\,\phi_c}\big)^2-1\big]^{1/2}}{n_1\,cos\,\phi_0-jn_0\big[ \big(\frac{sin\,\phi_0}{sin\,\phi_c}\big)^2-1\big]^{1/2}}.
\end{split}
\label{eq:Eq5}
\end{equation}

In Eq. (\ref{eq:Eq5}), \textit{$n_0$} represents the refractive index of vacuum (which is unity in our study) and \textit{$n_1$} represents the refractive index of the dielectric retro-reflector with intended dihedral angle offsets (which is 1.4607 for BK-7 as the intended material). The angle of reflection per TIR per ray is expressed as \textit{$\phi_0$} and the critical angle is expressed as \textit{$\phi_c$}. The subscripts \textit{s and p} represent the \textit{perpendicular-s} wave that is orthogonal to the plane of incidence, and the \textit{parallel-p} wave that resides within the plane of incidence. 

In our numerical simulation, we start with an array of equally spaced rays (such as an array of \textit{$256\times256$} rays with \textit{$0.4 mm$} spacing) at plane \textit{$z=70.72\,mm$} with specified polarization direction and make all rays propagate along \textit{$-z$} axis in the form of an incident planewave. Incident points are calculated per ray to determine its sequence of reflections, with its directions updated per TIR. In each TIR, the field strength of a ray is decomposed into s-wave and p-wave components, and Eq. (\ref{eq:Eq5}) is applied to update their complex field amplitude after each reflection, respectively. The end of the ray tracing is defined where the rays refract off the retro-vacuum interface after 3 TIRs and arrive at the reference plane of \textit{$z=70.72\, mm$}. Then the s-wave and p-wave components are converted into two independent wave-forms to derive their FFDPs respectively in wave optics. The phase profiles of the reflected s-wave and p-wave depend on the tracked phase changes \textit{$\phi$} per ray (diversified by the angle offsets) plus the relative phase change between the two orthogonal polarization direction (diversified by the depolarization). Irregular rays that miss the retro-reflector or don't go through 3 TIRs are labeled as invalid, and are removed from the FFDP calculations. 

\section{Simulation results}

Due to the lack of symmetry in the retro-reflector with dihedral angle offsets, it is not practical to use closed-form equations to derive the reflected waveform and find the optimized design. Instead, we rely on the numerical simulation to determine the optimized design. The intended angle offsets, under a coarse estimation based on the velocity difference between Earth and Moon, is approximated around \textit{$0.5\, arcsecond$}. Therefore, in both scenarios of REE and REN, we search over different modifications in \textit{$\Delta\Phi_{REE}$} and \textit{$\Delta\Phi_{REN}$} (shown in Fig. \ref{fig:Fig_2}) from \textit{0 arcsecond} to \textit{1.2 arcseconds} with increasing step size of \textit{0.04 arcsecond} to find the best case to improve photon returns for a primary sets of ground observatories capable of performing lunar laser ranging.  

\begin{figure}[ht!]
\centering
\includegraphics[width=0.6\linewidth]{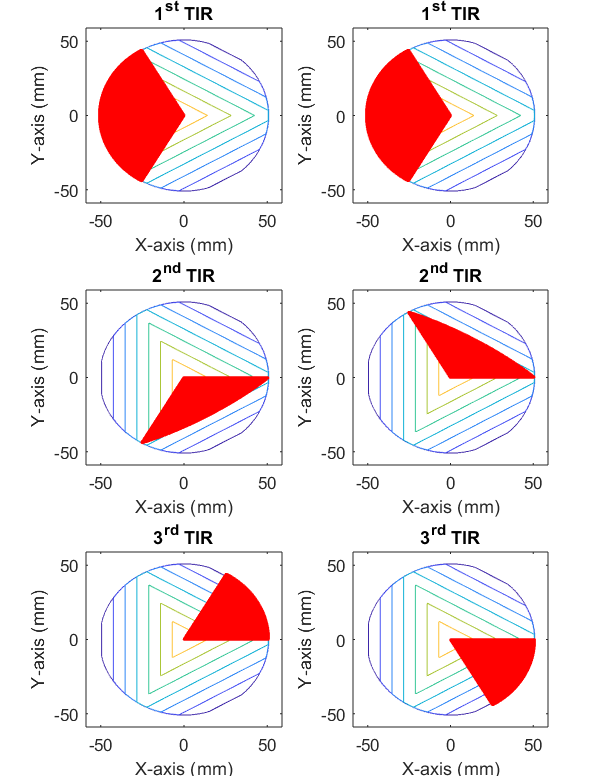}
\caption{Footprints of traced rays with different TIR sequences.}
\label{fig:Fig_3}
\end{figure}

We first show the sequence of beam reflections by footprints of rays on an unmodified retro-reflector in Fig. \ref{fig:Fig_3}, which serves as fundamental validation of the simulation approach. The colored contour lines show different depths of the 3 back faces of the retro-reflector in the 2D projection plots. Because of the 3-fold symmetry, we only show the REE orientation with first reflection happening on the one-third sector opposing to the alignment edge. The bifurcation happens at the \textit{$2^{nd}$} reflection where rays having \textit{$y<0$} are projected to the one-third sector below the x-axis, and rays having \textit{$y>0$} are cast to the one-third sector above the x-axis. Correspondingly in the \textit{$3^{rd}$} TIR, rays reflected off the remaining one-third sector leave the retro-reflector reversing the original incident direction and fill one-sixth of the aperture. It is easy to see that with the 3-fold symmetry, there are 6 different reflection sequences. It is also interesting to note that the \textit{$2^{nd}$} TIR doesn't involve the outer half areas in the back faces.

\begin{figure}[htbp!]
\centering
\includegraphics[width=0.7\linewidth]{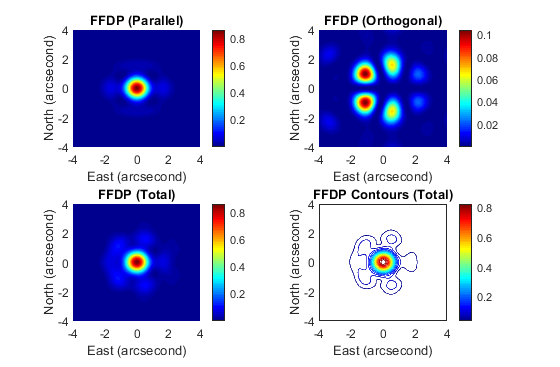}
\caption{Far field diffraction pattern of the 100 mm wide dielectric retro-reflector with its 2D projected real edge pointing East and incident plane wave East-West linearly polarized.}
\label{fig:Fig_4}
\end{figure}

Next we show agreement with established depolarization analysis of unmodified retro-reflectors for validation and baseline of showing the modification changes. Without loss of generality, we study the case of an input beam with East-West linear polarization (as commonly used in LLR experiments) under the REE orientation of the retro-reflector. The wavelength used is a \textit{$532\,nm$} green laser. And the depolarization results are shown in Fig. \ref{fig:Fig_4} that predict the FFDPs of the reflected beam. The intensity values are normalized by the central peak value at the center of the total FFDP. The FFDPs agree perfectly with well-established analysis done by Murphy \cite{murphy2013polarization}, Chang \cite{chang1971far} and Arnold \cite{arnold2002retroreflector}, so that the accuracy of the geometric-and-wave integrated simulation can be validated. It is easy to see that the ground stations (which typically intercept the return photons near 1 arcsecond) will inevitably get closer to the first zero of the reflected FFDPs in comparison with smaller retros used in previous Lunar ranging experiments. Therefore, it is of great interest to show how the angular offsets can mitigate the FFDPs for ground observatories in the REE and REN scenarios.

\begin{figure}[htb!]
\centering
\includegraphics[width=0.7\linewidth]{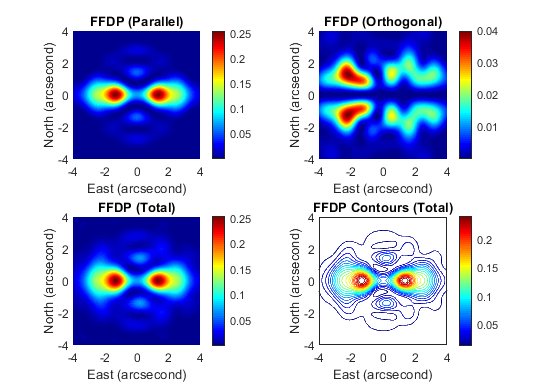}
\caption{Far field diffraction pattern of the modified dielectric retro-reflector with REE orientation, modification angle \textit{$\Delta\Phi_{REE}\!=\!0.5 \, arcsecond$} and East-West linear polarized input.}
\label{fig:Fig_5}
\end{figure}

\begin{figure}[htb!]
\centering
\includegraphics[width=0.7\linewidth]{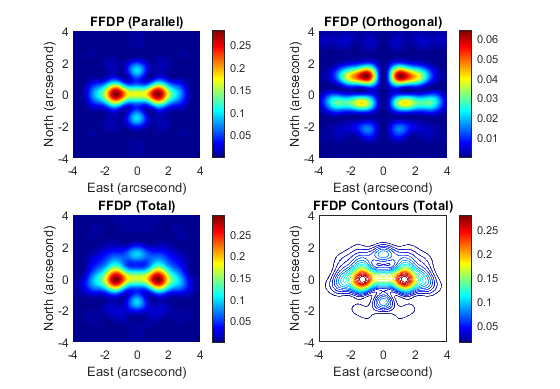}
\caption{Far field diffraction pattern from a modified dielectric retro-reflector with REN orientation, modification angle \textit{$\Delta\Phi_{REN}\!=\!0.5 \, arcsecond$} and East-West linear polarized input.}
\label{fig:Fig_6}
\end{figure}

As the Moon spins at the same speed as it orbits around the Earth except for trivial liberation, the reflection pattern can be understood as static per ranging shot. We show the simulation result with 0.5 arcsecond angular offset in \textit{$\Delta\Phi_{REE}$} for the REE orientation in Fig. \ref{fig:Fig_5}. And the simulation result with 0.5 arcsecond angular offset in \textit{$\Delta\Phi_{REN}$} for the REN orientation in Fig. \ref{fig:Fig_6}. Note that the offset strategies are different in the two scenarios to make the reflected patterns diverge in East-West directions, as shown in Fig. \ref{fig:Fig_2}. For consistency with Fig. \ref{fig:Fig_4}, we show both of the example cases with East-West linear polarization input and use the same normalization factor (the peak FFDP value in an unmodified retro-reflector) for the FFDPs shown in Figs. \ref{fig:Fig_5}-\ref{fig:Fig_6}. The results of North-South linear polarization input can be studied in the same way, which will not be elaborated here.

Both Fig. \ref{fig:Fig_5} and Fig. \ref{fig:Fig_6} show that the intended angle offsets can significantly lift photon returns for ground ranging observatories located near the \textit{1.0 arsecond} along the +x axis (pointing East). In fact, it can be observed that the majority of photon returns splits in two halves along the East/West directions after the modification. Therefore, it pivots the ground stations away from the situation of observing photon returns near the first diffraction zeros. Following the same fashion, we have estimated the improvement of photon returns per modification case for multiple observatories that are capable of performing Lunar laser ranging. The improvements are also shown by the ratios between actual photon returns per observatory per modification, and the peak photon return at the FFDP center for the same retro-reflector without modification.

In general, improvement varies from one observatory to another due to their differences in latitudes. In other words, due to the orbital motion of the Moon and the rotational motion of the Earth, the relative motion of the retro-reflector with regard to the Lunar Laser Observatory (LLO) on Earth is significant. For the previous Apollo arrays with small diameter (\textit{38 mm}) retro-reflectors, the loss for all LLOs are reasonably small (about \textit{$15\%$} when compared with the peak FFDP value), which do not need dihedral angle offsets. However, for the case of an unmodified 100 mm retro-reflector, the diffraction limited return beam will cause a loss over \textit{$80\%$} for certain LLOs. Therefore, it is important to make trade-offs and find a balanced design to increase ranging capacity for most available stations. The major ground observatories that are weighted in our study and design optimization are listed in Table \ref{tab:ground-observatories}.

\begin{table}[htb!]
\centering
\caption{\bf Major Lunar Ranging Stations Considered in Our Study}
\begin{tabular}{ccc}
\hline
location of observatory & latitude  & velocity aberration\\ &(degree) & (arcsecond)\\
\hline
Equator & $0$ & $0.769$ \\
Maui, Hawaii & $20$ & $0.809$ \\
Yunnan, China & $25$ & $0.830$\\
APOLLO, New Mexico & $32$ & $0.866$ \\
Materia, Italy & $40$ & $0.918$ \\
Grasse, France & $44$ & $0.949$ \\
Riga, Latvia & $57$ & $1.060$ \\
Ny Alesund, Norway & $79$ & $1.285$ \\
\hline
\end{tabular}
  \label{tab:ground-observatories}
\end{table}

\begin{figure}[tb!]
\centering
\includegraphics[width=0.70\linewidth]{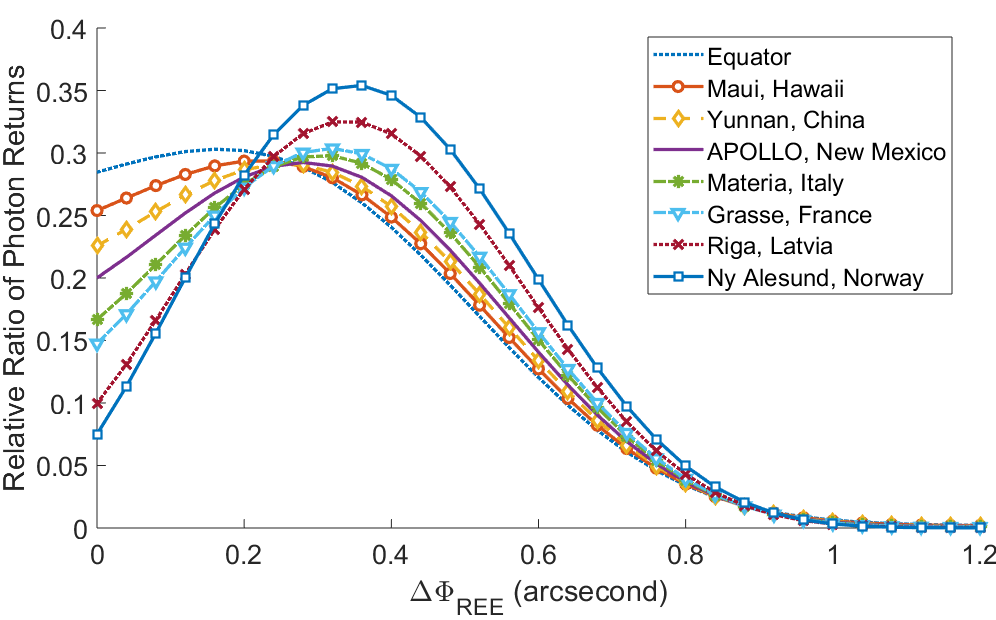}
\caption{Estimation of LLOs' photon return ratio (in relative scales) with the modified retro-reflector with REE.}
\label{fig:Fig_7}
\end{figure}

\begin{figure}[htp!]
\centering
\includegraphics[width=0.70\linewidth]{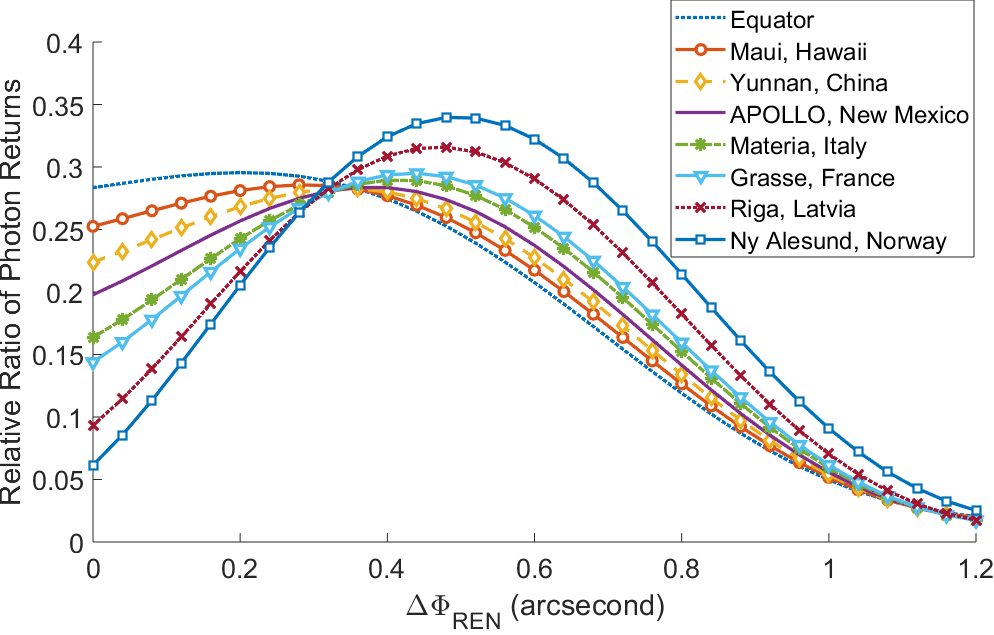}
\caption{Estimation of LLOs' photon return ratio (in relative scales) with the modified retro-reflector with REN.}
\label{fig:Fig_8}
\end{figure}

In Table \ref{tab:ground-observatories}, we have listed a number of primary LLOs at different latitudes to address the design issue of the modified retro-reflector. The relative velocity aberrations in terms of arcsecond for each LLO has been calculated and listed in Table \ref{tab:ground-observatories}, indicating the desirable angle that returns the offset beam back to the center of each LLO. With increasing levels of dihedral angle offsets, the relative ratios of photon returns for the listed major LLOs under the REE and REN scenarios are plotted in Fig. \ref{fig:Fig_7} and Fig. \ref{fig:Fig_8}, respectively.

In both Fig. \ref{fig:Fig_7} and Fig. \ref{fig:Fig_8}, we sweep through increasing levels of back angle offsets in the retro-reflectors to examine the photon returns at each LLO in Table \ref{tab:ground-observatories}. The intersection points on the vertical axis (y-axis) represent situations without the back-angle offsets. The \textit{"Relative Ratio of Photon Returns"} denotes that the values are normalized by the maximum photon return of an unmodified retro-reflector as the best case in photon return. It is evident to observe that all improvement curves are concave (uni-modal). In general, as the latitude of a LLO increases, more significant improvement can be witnessed with the offsets. And the optimized offset angle for a LLO increases at higher latitude.

\begin{figure}[htbp!]
\centering
\includegraphics[width=0.70\linewidth]{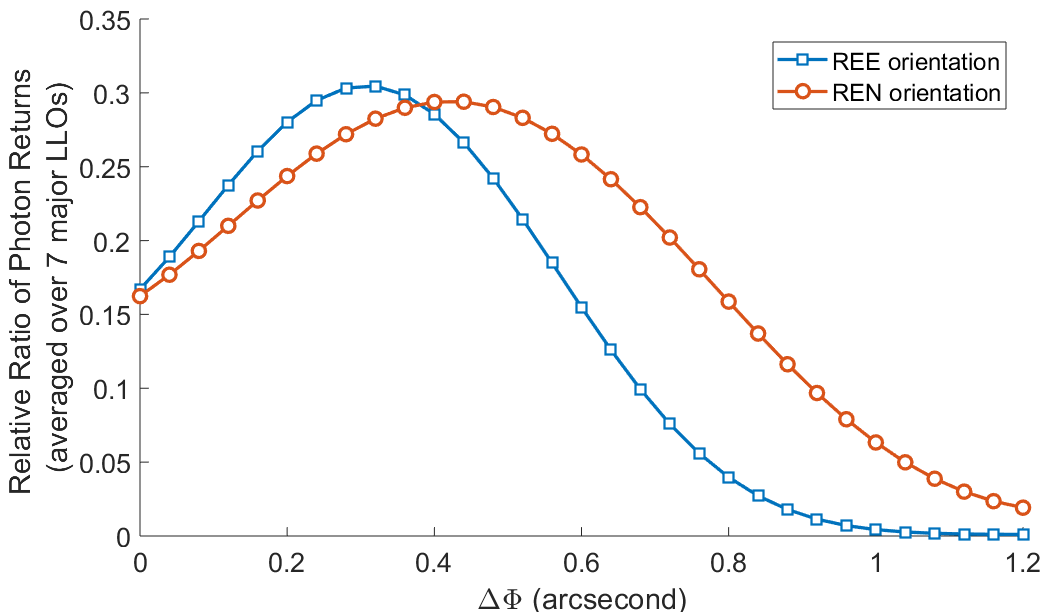}
\caption{Offset angle optimization based on averaged photon returns among 7 major LLOs.}
\label{fig:Fig_9}
\end{figure}

There are two optional criteria in determining the optimized angle offset of the retro-reflector, one is to maximize the average photon returns for all major LLOs (except the Equator), and the other is to provide all major LLOs with approximately the same photon returns. In the first criterion, the averaged photon return rate with different angle offsets are shown in Fig. \ref{fig:Fig_9}. As a result, the optimization is achieved at \textit{$\Delta\Phi_{REE}=0.3$} arcsecond for the REE case, and \textit{$\Delta\Phi_{REN}$=0.44} arcsecond for the REN case. When the second criteria is adopted, \textit{$\Delta\Phi_{REE}=0.24$} arcsecond gives the best uniformity for the REE case, and similarly \textit{$\Delta\Phi_{REN}=0.32$} arcsecond for the REN case. Given the facts that manufacturing tolerance is estimated around \textit{0.1 arcsecond} in adjusting the back angles, the exact optimization may not be achieved. Instead, the authors believe that if the actual angle offsets fall in range \textit{$(0.20, 0.50)$} arcsecond for the \textit{REE} scenario, and in range \textit{$(0.30, 0.60)$} arcsecond for the \textit{REN} scenario, good photon return rates can be harvested for all major LLOs based on Figs. \ref{fig:Fig_7}-\ref{fig:Fig_8}. 

Given that the peak photon return of a retro-reflector follows the power law of the $4^{th}$ order with regard to its diameter, the concluded design with intended dihedral angle offsets follows the same rule. In specific, more than \textit{$1/4$} of peak photon returns under ideally static Earth-and-Moon condition can be harvested for all LLOs through design optimization. The intended design avoids the situation where a LLO may suffer from low photon returns due to its uncompensated velocity aberration getting too close to the first zeros in the FFDP.  

\section{Evaluation of the first Manufactured NGLR}

\begin{figure}[ht!]
\centering
\includegraphics[width=0.4\linewidth]{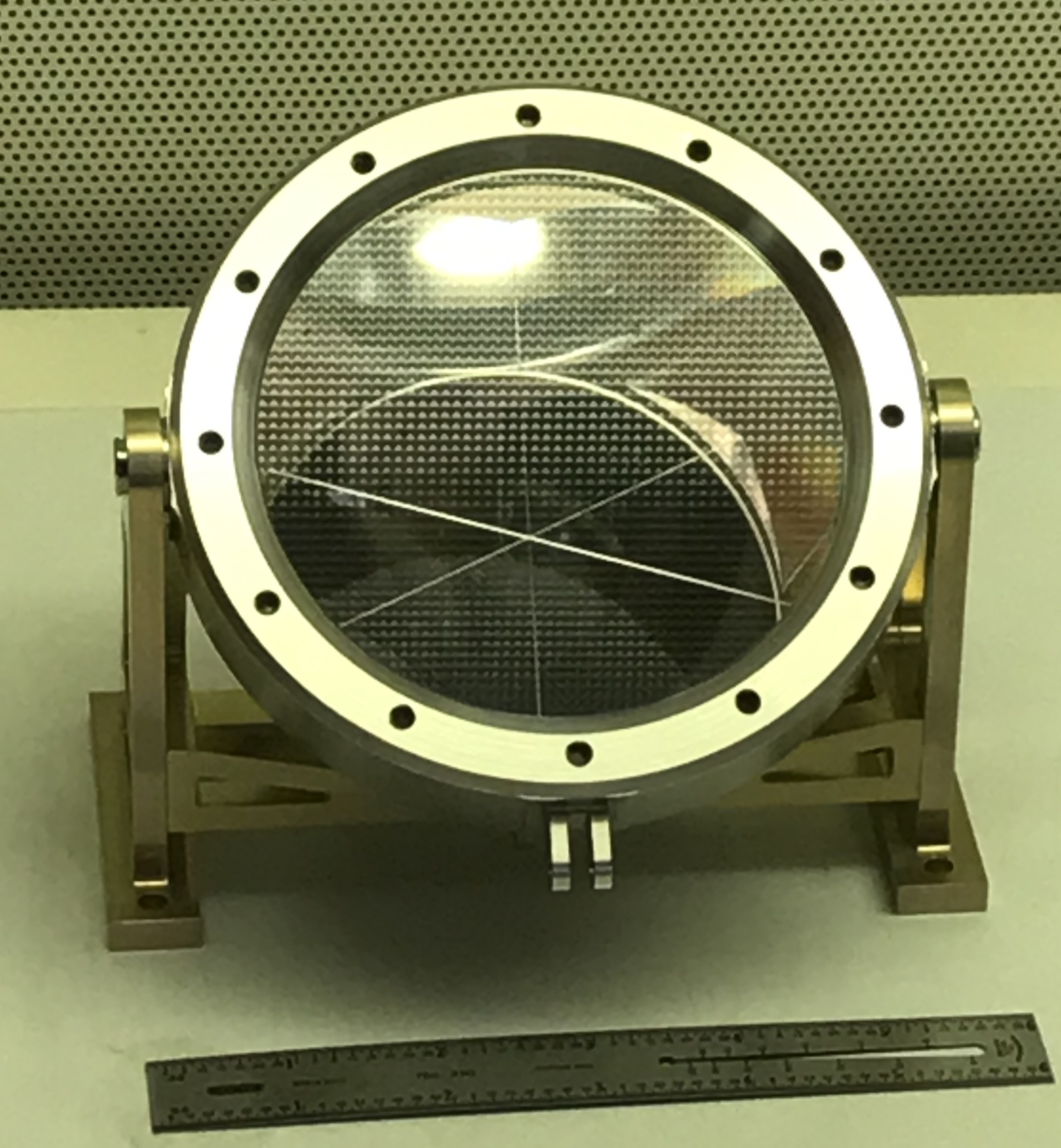}
\caption{Preliminary assembled view of the NGLR: UMD-02.}
\label{fig:NGLR}
\end{figure}

Manufacturing attempts have been made to implement a NGLR with the optimized design of \textit{$\Delta\Phi_{REE}$} in range \textit{$(0.24, 0.3)$} arcseconds. So that its effectiveness and challenges upon deployment can be better understood. Because there is no standard industrial technique for the back angle modification, post-process measurements by interferometers have been made to determine the three relative back angle offsets (RBAOs). The measured RBAOs are fed into the simulation for predicted FFDPs that can match with experimentally derived FFDPs using a wide flat beam as input to mimic situation on the moon. One notable challenge in the NGLR manufacturing is to maintain the irrelevant RBAOs as zeros, which is not achievable at current stage. Therefore, all three RBAOs must be measured with precision to predict its FFDP upon deployment. A view of the assembled NGLR is shown in Fig. \ref{fig:NGLR}.

We also find that the widely used single pass approach in interferometer measurement often leads to an error range near 0.1 arcsecond for the RBAOs. Such error is found through 120 degrees rotations of the NGLR where the outcomes are expected to be identical. The discrepancy is majorly resulted from the rotation asymmetry of the measurement beam employed by the interferometer. In other words, the circular asymmetry between NGLR reflected target beam (which is mirrored) and the reference beam leads to additional phase error, which adds to the RBAO measurements each time a 120 degrees of rotation is applied. Comparatively, the double pass approach that cancels the asymmetry by two reflections through NGLR provides much more stable results that fluctuate around 0.03 arcsecond or less. In evaluation, we rely on the double pass approach to determine accurate RBAOs. The zero-mean PEM per sector can also be added to the output beam to understand the influence of imperfect flatness of the NGLR's back faces.

The measured RBAOs results are shown in Table \ref{tab:three-retros}. The NGLR is given an ID of \textit{UMD-02}, followed by the measurement approach with \textit{SP} and \textit{DP} representing single pass and double pass, respectively. We use \textit{$\Delta\Phi_1$} for the RBAO of the east-pointing real edge. Evidently, both DP and SP indicate that the desirable RBAOs fall close to the correct range, while the remaining RBAO is not maintained at 0 arcsecond or close to that.  

\begin{table}[ht!]
\centering
\caption{\bf Interferometer Measured RBAOs of UMD-02 NGLR}
\begin{tabular}{ccccc}
\hline
Measurement id & Rotation & $\Delta\Phi_1$  &$\Delta\Phi_2$ & $\Delta\Phi_3$  \\ & (degrees) & (arcsecond) & (arcsecond) & (arcsecond) \\
\hline
UMD-02 SP & 0 & $0.2800$ & $-0.2725$ & $0.1750$ \\
UMD-02 DP-a & 0 & $0.3225$ & $-0.2378$ & $0.2763$ \\
UMD-02 DP-b & 120 & $0.2216$ & $0.3192$ & $-0.2834$ \\
UMD-02 DP-c & 240 & $-0.2852$ & $0.2023$ & $0.3212$ \\
\hline
\end{tabular}
  \label{tab:three-retros}
\end{table}

\begin{figure}[ht!]
\centering
\includegraphics[width=0.97\linewidth]{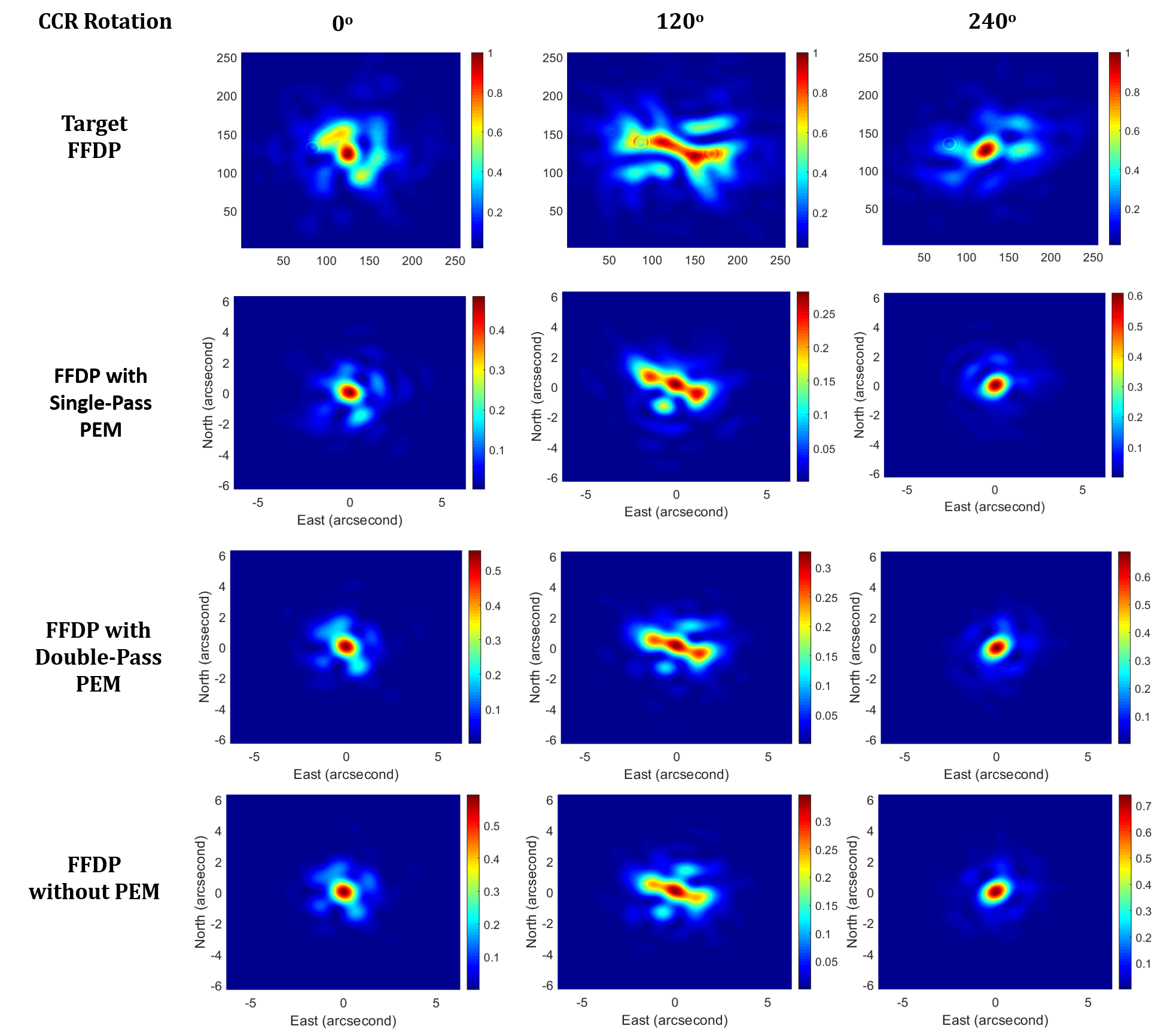}
\caption{Matching Experimental FFDPs with interferometer inferred FFDPs.}
\label{fig:FFDP_match}
\end{figure}

In Table \ref{tab:three-retros}, the first measurement is performed by Precision-Optics using a Wyko 6000 interferometer with the single pass configuration. Due to the aforementioned precision concern, the second measurement has been performed with the double pass configuration, which is provided by 4D technology using an AccuFiz interferometer. The measurements under different configurations essentially agree with each other except that the \textit{DP} configuration renders a much smaller variance through multiple rotations of 120 degrees. Therefore, we pick the averaged values of the 3 DP measurements to determine its RBAOs as $[0.3210,-0.2688,0.2334]$ arcseconds, and use them for our evaluation. To further validate the accuracy of the measured RBAOs, experimental FFDPs are obtained at NASA Goddard Space Flight Center (GSFC) for comparisons. In the experiment, the reflection pattern from a wide collimated laser beam input (with a diameter >100 mm and wavelength of 532 nm) is measured by a beam profiler (Newport: LBP2-HR-VIS2) at the back focal plane of a lens system. The experimental beam has a linear polarization along -45 degrees while the NGLR has its first real edge $\Delta\Phi_1$ orthogonally aligned with the beam polarization at 0 degrees. By the 3-fold rotations of 120 degrees, 3 distinctive FFDPs have been obtained. The fundamental matching between the experimental FFDPs and the simulated FFDPs using the RBAOs are shown in Fig. \ref{fig:FFDP_match}. 

\begin{figure}[htbp!]
\centering
\includegraphics[width=0.75\linewidth]{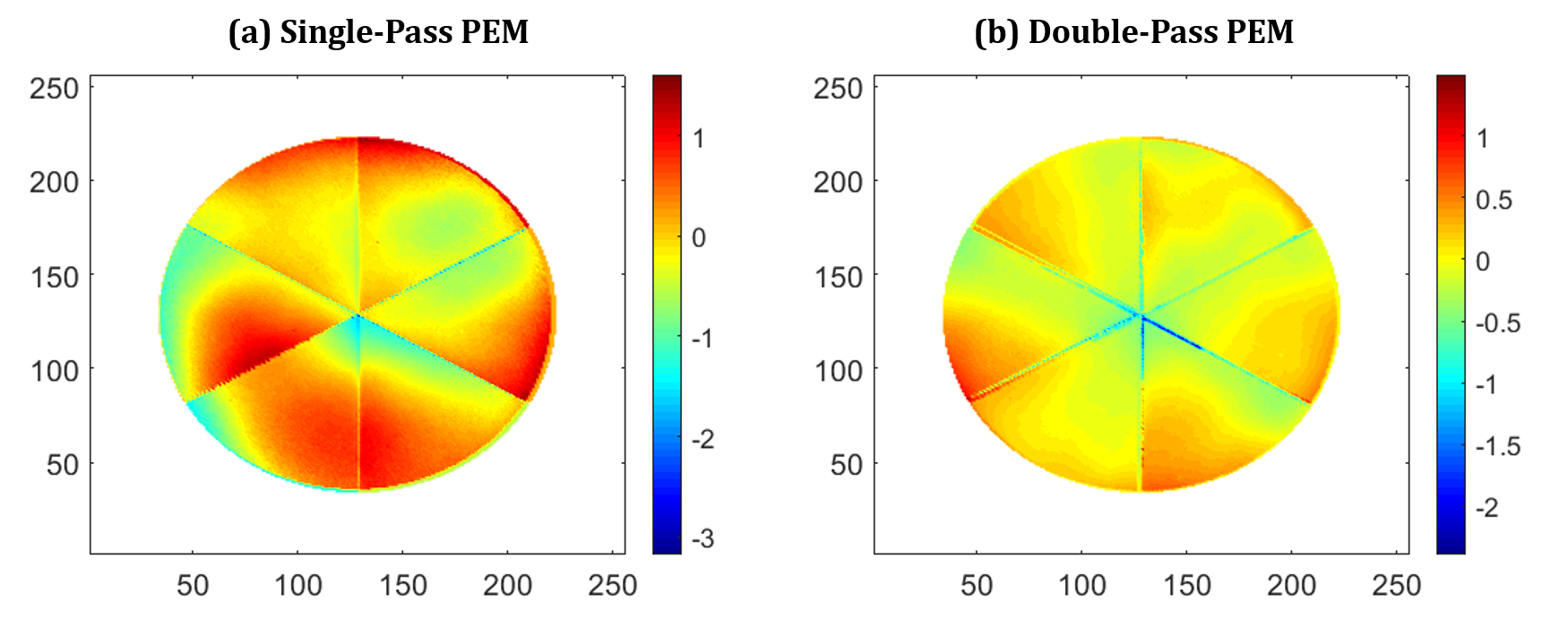}
\caption{Interferometer measured PEMs with SP and DP configurations.}
\label{fig:PEMs}
\end{figure}

In Fig. \ref{fig:FFDP_match}, basic agreements can be found between independent interferometer and beam profiler measurements. The first row of figures in Fig. \ref{fig:FFDP_match} represent the experimental FFDPs at GSFC with axes representing pixel indices. The second row of figures present simulated FFDPs by using the DP configured RBAOs and the SP configured PEM, with axes representing the actual angular space in diffraction. Comparatively, the third row of figures present simulated FFDPs by using the DP configured RBAOs and the DP configured PEM. And the fourth row represents FFDPs assuming zero PEM (pefect flatness in the back faces). The similarity in the fundamental shapes of the FFDPs per rotation show that the interferometer measured RBAOs are reasonably accurate. There are certain levels of mismatches between the beam profiler and the interferometer approaches, yet the deviation is significant enough to revert the simulation predictions. Therefore we can rely on the simulation to predict other FFDPs under different rotations of the NGLR as well as beam polarization. The SP configured PEM is obtained at 0 degrees device rotation at 4D technology. The DP configured PEM only provides half aperture PEM per measurement, and is assembled by adding all 3 DP measurements shown in Table \ref{tab:three-retros} and dividing the overlapped sectors by half. A view of the SP and DP retrieved PEMs by the interferometer is shown in Fig. \ref{fig:PEMs}. The 2 PEMs are weakly correlated with a correlation coefficient of 0.1521. The RMS of the SP configured PEM is $\frac{1}{12}\lambda$, while the RMS of the DP configured PEM is $\frac{1}{27}\lambda$. At such level, the PEM only causes minor changes to the FFDPs when compared with the influences of input beam polarization and NGRL rotation. The same can be witnessed in Fig. \ref{fig:FFDP_match}, where the FFDPs do not have significant changes with and without the PEM provided by the interferometer. As the interferometer provided PEMs do not provide conclusive explanation for the residue differences between the $1^{st}$ (GSFC experimental FFDPs) and the $4^{th}$ row (RBAO inferred FFDPs) in Fig. \ref{fig:FFDP_match}, we do not include them for the current evaluation of the NGLR.   

\begin{figure}[htbp!]
\centering
\includegraphics[width=0.60\linewidth]{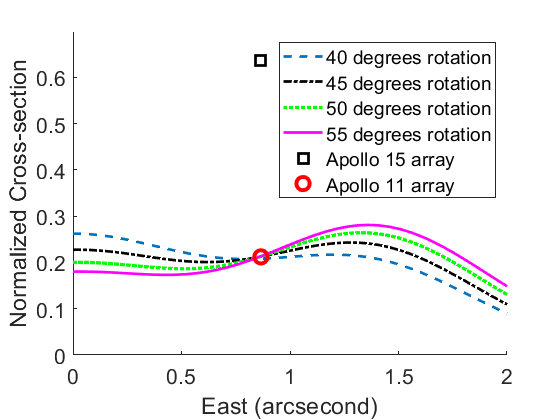}
\caption{Evaluation of the manufactured UMD-02 NGLR with hypothetical deployment.}
\label{fig:PhotonReturnPredict532}
\end{figure}

Based on the above validation and interferometer data, we predict photon returns by the UMD-02 NGLR for the common LLR beam setup with East-West linear polarization and wavelength of 532 nm. The result is shown in Fig. \ref{fig:PhotonReturnPredict532}.

In Fig. \ref{fig:PhotonReturnPredict532}, normalized cross-section values have been used based on conventions in evaluating retro-reflectors \cite{degnan2012tutorial,paris2019comparison,arnold2002retroreflector}. Cross section values for the previous Apollo 11 and Apollo 15 arrays have been calculated and co-plotted for comparison, which a special focus near the APOLLO LLO. As the Apollo arrays have low diffraction loss (around $15\%$) for all LLROs, their values are maintained near the same level for other LLOs. All the values are normalized based on the peak cross-section \textit{$\sigma_{cc}=2.54 \times 10^{9} m^{2}$} of an unmodified solid retro-reflector with 100 mm diameter and assuming that an unified wavelength of 532 nm has been used and the reflection is lossless. Therefore, the normalized cross-section values for the UMD-02 NGLR are the same as the relative photon returns as discussed in the previous sections. Because the UMD-02 has not accurately implemented the optimized RBAOs, exhaustive simulation searches have been performed to find best rotations to return the photons to the East axis. In Fig. \ref{fig:PhotonReturnPredict532}, rotations with 40, 45, 50 and 55 degrees are shown simultaneously (where 0 degree rotation has the real edge with the 0.3210 arcsecond RBAO pointed East) to reflect changes near the East 1.0 arcsecond locations (where most of the LLOs are "located"). It can be concluded that if the UMD-02 NGLR were to be deployed, its photon return rate is approximately equal to the Apollo 11 array containing 100 retro-reflectors that have a diameter of 38 mm. The best deployment angle of the UMD-02 is 45 degrees (counter-clockwise) for stationary returns over all LLOs. A tolerance of $\pm5$ degrees deviation can be given where the returns are still balanced for all LLOs. When compared with the APOLLO 15 retro array (which uses 300 dielectric retro-reflectors of 38 mm aperture diameter), the average ratio is approximately 0.4. In fact, based on theoretical estimation, the new retro-reflector is expected to achieve \textit{$48/300=0.16$} in ratio to the APOLLO 15 retro array based on gains in larger aperture (48:1) and losses in retro-unit counts (1:300). Additional considerations such as the actual locations of APOLLO LLO in the FFDPs (a ratio of \textit{$0.2 : 0.75$}) and estimated degeneration (dust accumulation) of the APOLLO 15 retro array (with an optimistic factor of 10) lead to a boost factor of 2.7. As a result, the above gives 0.43 for the photon return ratio between the manufactured UMD-02 NGLR and the Apollo 15 array. This means without requesting the LLOs to change the common configurations of the ranging lasers for better returns, the current NGLR product is already comparable to previous Apollo arrays that succeed in LLR and are still functioning. Given comparable photon returns, the NGLR wins by confining the return photons within a single retro-reflector, so that the ranging error caused by a tilt angle of the Apollo arrays facing Earth and the uncertainty of which unit returns the received photon will be minimized (to a value much smaller than 1.0 mm). The light-weight ($\approx$1.5 kg per package) and compactness of the NGLR are also beneficial for space missions to carry such device. With future advancement in NGLR manufacturing, as well as investigation on the customized beam protocols and implementation for each LLO based on its latitude, LLR with much improved accuracy and reliability can be achieved.

\section{Evaluation of the second Manufactured NGLR}
Based on the evaluation of the first NGLR (UMD-02), we found that the baseline of NGLR performance can be largely predicted by the DP interferometer measurements of the NGLR. The second NGLR (UMD-03) has been manufactured in 2020 by ZYGO Corporation with newer technology to ensure surface flatness. The second NGLR is designed based on the optimization of REN configuration with \textit{$\Delta\Phi_{REN}$=0.32} arcsecond that fits the common use of ranging pulses with East-West linear polarization. The measured angles are shown in Table \ref{tab:new-retro}.

\begin{table}[ht!]
\centering
\caption{\bf Interferometer Measured RBAOs of UMD-03 NGLR}
\begin{tabular}{ccccc}
\hline
Measurement id & Rotation & $\Delta\Phi_1$  &$\Delta\Phi_2$ & $\Delta\Phi_3$  \\ & (degrees) & (arcsecond) & (arcsecond) & (arcsecond) \\
\hline
UMD-03 DP-a & 0 & $0.3175$ & $0.0513$ & $-0.04$ \\
UMD-03 DP-b & 120 & $-0.051$ & $0.3431$ & $0.0382$ \\
UMD-03 DP-c & 240 & $0.047$ & $-0.033$ & $0.3337$ \\
\hline
\end{tabular}
  \label{tab:new-retro}
\end{table}

\begin{figure}[htbp!]
\centering
\includegraphics[width=0.70\linewidth]{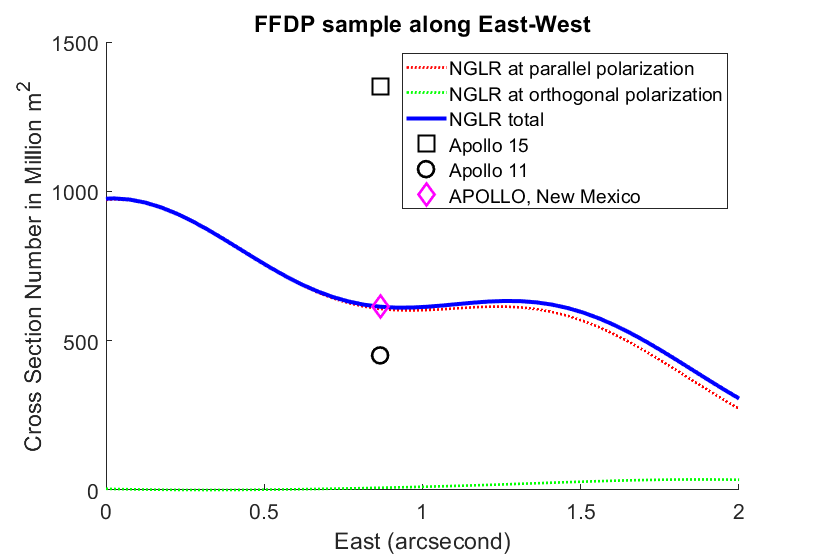}
\caption{Evaluation of the manufactured UMD-03 NGLR with REN orientation in deployment.}
\label{fig:PhotonReturnPredict532new}
\end{figure}

When compared with the UMD-02 NGLR, the UMD-03 NGLR is a better match with the optimization purpose. In particular, the new manufacturing process is able to maintain the other two RBAOs close to 0 arcsecond. Due to the good symmetry of the product, the best angle for deployment follows the principle of having the real edge with the largest RBAO pointing North. By combining the three DP measurements, the set of RBAOs for the UMD-03 NGLR are evaluated as (0.3314, 0.0455, -0.0413) arcseconds, with predicted return (measured by the cross sections in million meter squares) shown in Fig. \ref{fig:PhotonReturnPredict532new}.

In Fig.\ref{fig:PhotonReturnPredict532new}, the cross section calculation for the APOLLO 15 array agrees with Degnan's notation and the tabulated ILRS value of $1.4\times10^9 m^2$ \cite{degnan2012tutorial}. The results predict that UMD-03 NGLR will have $20\%$ more returns than the APOLLO 11 array, which is suitable for actual deployment.  

\section{Conclusions and discussions}
In this work, we have systematically studied the optical design of using intended dihedral angle offsets for the NGLRs to be deployed on the Moon, which solves the problem of low far-field diffraction photon returns for large aperture solid retro-reflectors, and will significantly improve the ranging accuracy by confining all the returned photons in the same retro-reflector to minimize their path differences in reflection. In specific, the purpose of the offsets, the numerical simulation approach, the consequences of the modification with regard to ground LLOs, as well as one manufactured NGLR have been discussed with extensive details. We point out through combined simulation and experimental measurements that the current UMD-02 NGLR (as a non-optimized unit) is already suitable for LLR with high accuracy and reasonable photon returns.

\begin{figure}[ht!]
\centering
\includegraphics[width=0.70\linewidth]{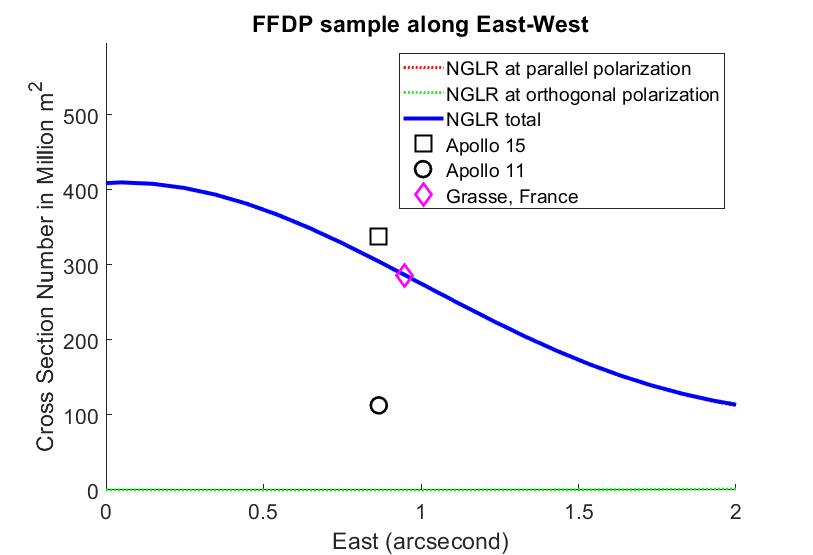}
\caption{Evaluation of UMD-03 NGLR with East-West linear polarization input and 1064 nm wavelength for the ranging laser-pulses.}
\label{fig:PhotonReturnPredict1064new}
\end{figure}

Recently, wavelength of 1064 nm have also been used for LLR \cite{courde2017lunar}. The corresponding returns for the UMD-03 NGLR is evaluated in Fig. \ref{fig:PhotonReturnPredict1064new}. The situation for the Grasse laser station (France) will improve more when working with the UMD-03 NGLR. 

Our future study will address additional issues such as heat and radiation analysis and management of the NGLR housing over the approximated 250K temperature sweep over a lunar day, and mechanisms of dust precaution upon and after deployment to better understand and predict the NGLRs' quality over long terms of uses in future LLR experiments. Due to the use of the same dielectric material and device geometry, the study is majorly conducted through embedded sensors in the UMD-01 NGLR in a vacuum chamber (that is transferable to the study of the UMD-02 and UMD-03 NGLRs).  

In a wide sense, ranging and localizing space vehicles and celestial bodies with velocity aberrations and astronomical distances (typically above 1 light second) will become a common task in numerous space applications. Due to the similarity among these problems, the current study and technique designed for the NGLR provides valuable guidance for future retro-reflectors to be used for deep-space missions.   

\section*{Funding Information}
The work is sponsored through NASA's Commercial Lunar Payload Services (CLPS) project.

\section*{Acknowledgement}
The authors sincerely thank Dr. Stephen Martinek at 4D Technology for providing accurate interferometer measurement of the UMD-01 NGLR. We thank NASA's Goddard Space Flight Center (GSFC) for supporting the experimental measurements of the UMD-02 NGLR's FFDPs. We thank Dr. Richard Boland from Zygo Corporation for his kind technical support and manufacturing of the UMD-03 NGLR. We also thank Dr. Giovanni Delle Monache at National Institute for Nuclear Physics (INFN, Italy) for the initial manufacturing and support of the UMD NGLRs. 

\noindent\textbf{Disclosures.} The authors declare no conflicts of interest.

\bibliographystyle{unsrt}
\bibliography{CCR} 

\end{document}